\documentclass[twocolumn,secnumarabic,amssymb, nobibnotes, aps, prd]{revtex4-1}
\usepackage{graphicx}  

\usepackage{color}

\begin{document}

\title{
Hidden orders in amorphous structures: extraction of nearest neighbor networks of amorphous Nd-Fe alloys with Gabriel graph analyses
}
\author{Asako Terasawa}
\email{terasawa.a.aa@m.titech.ac.jp}
\author{Yoshihiro Gohda}
\affiliation{
Department of Materials Science and Engineering, Tokyo Institute of Technology,
Nagatsuta-cho 4259, Midori-ku, Yokohama 226-8502, Japan}
\date{1 June 2018}%
\begin{abstract}
Using the scheme of 
Delaunay and 
Gabriel graphs, we analyzed the amorphous structures of computationally created Nd-Fe alloys for several composition ratios based on melt quench simulations with finite temperature first-principles molecular dynamics.
By the comparison of the radial distribution functions of the whole system and those derived from the Delaunay and Gabriel graphs, 
it was shown that the Gabriel graphs represent the first nearest neighbor networks well in the examined amorphous systems.
From the Gabriel graph analyses, we examined the coordination structures of amorphous Nd-Fe alloys statistically.
We found that the ranges of distributions of coordination numbers vary depending on the composition ratio.
The angular distributions among three adjacent atoms were also analyzed, and it was found that the angular distributions behave differently in the Nd-rich and Fe-rich samples.
We found that the orders in the amorphous system becomes 
stronger as increasing the Nd ratio, which corresponds to the appearance of crystalline grain boundary phases at high Nd composition ratio 
[T.~T.~Sasaki \textit{et al.}, Acta Mater. \textbf{115}, 269--277 (2016)].
\end{abstract}

\maketitle

\section{Introduction}\label{sec:intro}

Amorphous materials have much significance recently in applications of electrostatic and magnetic materials.
For example, amorphous metal-oxide semiconductors have been investigated for many ways of electronic device applications \cite{elect1,elect2,elect3}, and amorphous magnetic alloys have been investigated as good candidates for soft magnetic materials \cite{soft1,soft2,soft3,soft4,soft5,soft6,soft7}.
Moreover, amorphous phases in microstructures attract much interest in the studies of permanent magnets, because the microstructures in permanent magnets, such as the grain boundary (GB) phases, have strong effects on the performance of permanent magnets \cite{GB1,GB2,GB3,GB4,GB5,GB6}.
Particularly, an experiment revealed the relationship among the crystallinity of GB phases, their detailed arrangements and compositions \cite{GB_amor}.
In the experiment, it was found that the amorphous GB phases appear at the interfaces perpendicular to the $ab$ planes of Nd$_2$Fe$_{14}$B grains with the Nd composition ratio of about 40~\%, and the crystalline GB phases appear at the interfaces perpendicular to the $c$ planes of Nd$_2$Fe$_{14}$B grains with the Nd composition ratio of about 60~\%. 
In comparison, theories of electronic and magnetic properties of amorphous materials are in early stage and theoretical interpretations of the electronics and magnetisms of amorphous materials are still difficult and challenging tasks.

For many decades, the structural properties of amorphous systems have been widely studied \cite{amor1,amor2,amor3,amor4,amor5} in terms of statistical approaches such as classical molecular dynamics (MD).
Out of those studies, many characteristic functions and parameters (e.~g.\ radial distribution functions, angular distribution functions and bond orientational order parameters) have been calculated as statistical values of amorphous systems to describe the structural properties of amorphous systems \cite{amor6}.

On the other hand, quantum mechanical treatments have been adopted to understand electron-related phenomena such as electron conduction and magnetism for various materials from the atomistic point of view.
In particular, the density-functional theory has become a standard in this field because of its capability to treat relatively large systems with relatively high accuracy of multi-electron treatment.
Various studies have also been made to adopt statistical treatments into density-functional approaches, such as melt-quench first-principles MD simulations which have been succeeded to give accurate structural properties of amorphous systems \cite{MQ_FP,MQ_FP2,MQ_FP3,MQ_FP4}, and the calculated amorphous systems scaled up to 576 atoms for the times steps of 14,000 in the recent density functionals study of metal-sulfide amorphous materials \cite{SO}.

In spite of those attempts, however, there is a large gap between the understandings of electron-related properties and the structural understanding of amorphous systems. The problem here is that, amorphous structures are determined only statistically whereas electron-related properties are calculated from a particular structure.
It is not safe enough to discuss the universal features by a single structure calculation because such a calculation does not necessarily reflect the statistics in the amorphous systems well. To take the statistical averages of detailed electron-related properties for a long time scale is still a hard task for first-principles based computational techniques, and may make the relationship between the electron-related properties and the structural characters ambiguous.

Therefore, our purpose is to depict the structural characters of amorphous systems which can be determined uniquely for a particular structure and which has a correspondence with the statistical depiction, and to apply it to the analyses of first-principles calculations to find the relationships between structural and electronic characters.
A similar mathematical approach to represent amorphous structures had previously been done
by Nishio \textit{et al.}, by performing Voronoi tessellation regarding the atomic positions \cite{Voronoi1}.
The Voronoi tessellations, however, often result in the complicated structures of which variations are named by the sets of polyhedral tilings \cite{Voronoi2,Voronoi3},
which may not give a straightforward relationship with electronic characters.

\begin{figure}[th]
\begin{center}
\includegraphics[width=75mm]{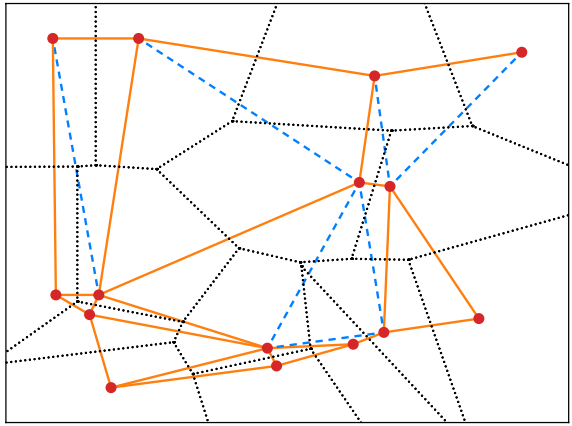}\\
\end{center}
\caption{
A Relationship of Voronoi polyhedra and Delaunay and Gabriel graphs for a given set of points in a two-dimensional system.
The black dotted lines represents the Voronoi polyhedra, which are consisting of the perpendicular bisectors for pairs of adjacent points.
The solid and dashed lines represent the edges of Delaunay graph, which represents the contact of Voronoi polyhedra belonging to the corresponding points.
The orange solid lines represent the edges of Gabriel graph, where some edges of the Delaunay graph are disconnected when there are intercepting Voronoi region belonging to other atoms. The blue dashed lines represent the edges that are included to the Delaunay graph but excluded from the Gabriel graph.
\label{fig:scheme}
}
\end{figure}

In this paper, we propose a simple method to depict nearest neighbor (NN) networks in amorphous systems based on
the graph theory. We compared two types of graphs, the Delauney graph \cite{Delaunay} and the Gabriel graph \cite{Gabriel}.
The Delaunay triangulation is one of a standard methods to represent neighboring structures in disordered systems, and this method has a dual correspondence with the Voronoi tessellation \cite{Delaunay2}. 
The Gabriel graph is also a representation of neighboring networks in disordered systems, and the Gabriel graph constructs a partial graph of the Delaunay graph for a given set of nodes. Figure \ref{fig:scheme} show the relationship among Voronoi polyhedra, Delaunay and Gabriel graphs for a two-dimensional system as a schematic. 

Using these two types of graphs, we represented the nearest neighbor structures in Nd-Fe alloy created computationally using melt-quench molecular dynamics and first-principles calculation.
To determine which graph gives better expression of NN sites in Nd-Fe amorphous alloy we compared the radial distribution functions (RDFs) of the whole system and the RDFs of the edges of Delaunay and Gabriel graphs.
it is revealed that the Gabriel graphs give a good depiction of first NN sites, whereas the Delaunay graphs overestimate the number of first NN sites.

We also performed the Gabriel graph analyses to the Nd-Fe alloys created by first-principles calculations, and found that the Gabriel graphs depict the first NN pairs of atoms well in the examined systems.
We made detailed analyses of the Gabriel-graph-based coordination structures of amorphous Nd-Fe alloys statistically for the results of finite-temperature first-principles MD simulations. We found that the coordination numbers of atoms distribute in certain ranges depending on the composition ratio of Nd-Fe alloy. We also analyzed the angular distributions for pairs of adjacent edges in the Gabriel graphs, and found that the features of angular distributions differ significantly in the Fe-rich cases and in the Nd-rich cases.
It is found that the ordered structures in the angular distribution functions vary depending on the Nd composition ratio. This can be related to
the difference in crystallinity of 
GB phases in the experimental result~\cite{GB_amor}.

\section{Computational method}

For the first-principles calculations, we used OpenMX code \cite{OpenMX}. In the framework of OpenMX, we adopted the Perdew-Burke-Ernzerhof exchange-correlation functional \cite{GGA-PBE} with the generalized gradient approximation. For the pseudoatomic orbital basis sets, we adopted $s1p1d1$ basis set for Fe and the $s2p1d1$ basis set for Nd, where 3$p$ states of Fe and $5s$ and $5p$ states of Nd are treated explicitly as valence states. Cutoff radii were set as 6.0 Bohr for Fe and 8.0 Bohr for Nd. We adopted the fully relativistic pseudopotentials generated by the Morrison-Bylander-Kleinman scheme \cite{MBK}, with the $4f$ state of Nd taken into account as a spin-polarized core state by considering occupation of three electrons. We used $1\times 1\times 1$ $k$-grid with the cutoff energy of 300 Ry.

The convergence criteria for the force and the total energy were chosen as 1.0$\times$10$^{-3}$ Hartree/Bohr and $10^{-6}$ Hartree, respectively, and the time step $\Delta t$ was set as 1 fs. For the finite temperature MD simulations, we adopted the velocity scaling method.

\section{Melt-quench molecular dynamics simulation}

The amorphous structures were prepared by the melt-quench method \cite{MQ1} based on first-principles molecular dynamics as follows:
A bcc-like periodic alloy having 27 atoms of Fe and 27 atoms of Nd was prepared for the first step.
Second, a certain number of atoms were substituted for desired compositions. 
We prepared 7 systems having composition ratios of Nd $= 20~\%,~31~\%,~41~\%,~50~\%,~59~\%,~69~\%$ and 80~\%. 
Alloys were melted at 4000 K for 1 ps, quenched to 300 K in 2 ps and then stabilized at 300 K for 2 ps by first-principles molecular dynamics (MD).
A structural optimization regarding the internal coordinates and the lattice vectors was applied after the melt-quench procedure for each composition.
Then, the systems were annealed at 900 K for 1 ps, cooled to 300 K in 2 ps and then stabilized at 300 K for 2 ps, and second structural optimizations followed afterwards.
And then the procedures of melting-quenching-stabilizing, annealing-cooling-stabilizing, and structural optimization are applied repeatedly to obtain 6 independent samples for each composition ratio.

\begin{figure}[th]
\begin{center}
\includegraphics[width=6.5 cm]{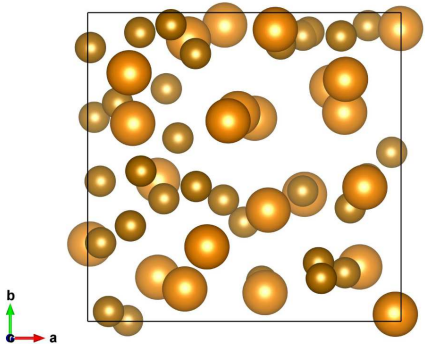}
\end{center}
\caption{
An example of calculated amorphous Nd-Fe alloy, which contains 22 atoms of Nd and 32 atoms of Fe, and its composition ratio is Nd~=~41~\%. In the figure, the large spheres correspond to the Nd atoms and the small spheres correspond to the Fe atoms.
}
\label{fig:sample}
\end{figure}

Figure \ref{fig:sample} shows an example of the atomic coordinates of Nd-Fe alloy with the composition ratio of Nd $= 41~\%$.
It is possible to see a disordered structure in in Fig.~\ref{fig:sample}.

For each sample created by the procedures above, the finite temperature first-principles MD simulation at 300 K was performed for 4 ps.
We performed statistical analyses explained below for the coordinates of these MD calculation results.
First, we calculated the RDFs as in the following formula:
\begin{equation}
g_{\mathrm{AB}}(r)
=\frac{1}{4\pi r^2}\frac{1}{N_{\mathrm{time}}}\sum_{n=1}^{N_{\mathrm{time}}}\sum_{i\in\mathrm{A}}\sum_{j\in\mathrm{B},j\ne i}
\!\!\!\delta\left(r-R_{ij}(n\Delta t)\right),
\label{eq:RDF}
\end{equation}
where A and B represent atomic species, $i\in\mathrm{A}$ means that atom $i$ belongs to atomic species A, $R_{ij}(n\Delta t)$ corresponds to the distance between atoms $i$ and $j$ at time $t=n\Delta t$ and $n$ runs from $1$ to $N_{\mathrm{time}}=4000$. In this paper, we used an approximated delta function as $\delta(r) \approx (\sqrt{2\pi}\sigma)^{-1}\exp\left[ -r^{2}/(2\sigma ^{2})\right]$ with $\sigma=1.0\times 10^{-2}~[\mathrm{\AA}]$.
We took into account all the atoms in the unit cell for summation for atom $i$, and summation over atoms in different cells were considered for atom $j$.
Regarding the results in the following sections, we call the RDFs by this definition as the ``whole'' RDFs when necessary.

\begin{figure}
\begin{center}
\includegraphics[width=65mm]{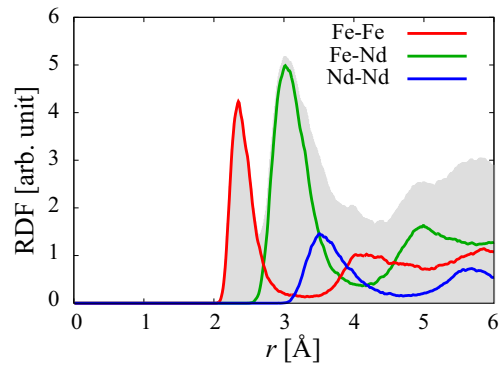}
\end{center}
\caption{
Radial distribution functions for the Nd-Fe alloy at the composition ratio of Nd $= 41~\%$.
The red, green and blue solid lines corresponds to the RDF components of Fe-Fe, Fe-Nd and Nd-Nd, and gray area corresponds to the sum of all the components.
All the functions are the averages of 6 independent samples.
\label{fig:RDF}
}
\end{figure}

As an example, we show the RDFs at Nd = 41~\% in Fig.~\ref{fig:RDF}.
The lines in Fig.~\ref{fig:RDF}, corresponding to the RDF components of different element pairs, show steep rises from the smallest value of $r$, drop sharply and then show broad peak structures for larger $r$.
These features are typical to amorphous systems \cite{amor}.

\section{Graph analysis for amorphous systems: the Delaunay and Gabriel graphs}

\begin{figure}[th]
\begin{center}
\includegraphics[width=75mm]{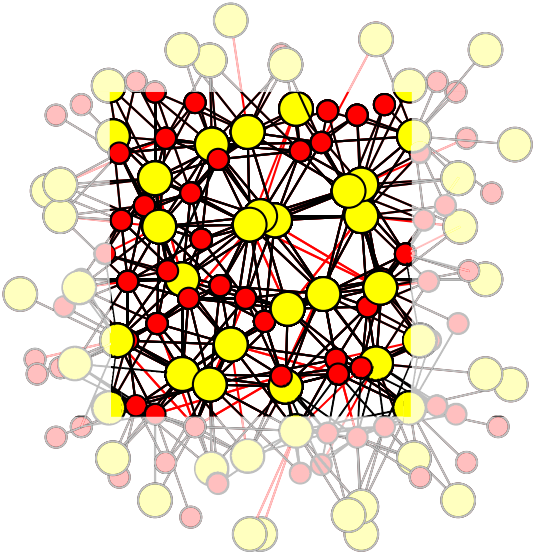}
\end{center}
\caption{
Delaunay and Gabriel graphs derived from the amorphous Nd-Fe alloy which contains 41~\% of Nd.
The large yellow circles and small red circles correspond to Nd and Fe atoms, respectively.
The black lines represent the edges of Gabriel graph and the red lines show the edges that are included in the Delaunay graph but excluded from the Gabriel graph.
The pale region in the figure represents the outside of a unit cell.
\label{fig:ggraph_example}
}
\end{figure}

Our purpose is to find the unified depiction of structural character that is applicable to a particular set of atomic coordinates.
For this purpose, we examined two types of graphs; the Delaunay and Gabriel graphs.
The criteria to put the edges in the Delaunay and Gabriel graph for a given atomic coordinates can be defined as follows.
For the Delaunay graph, we need to find at least one point $\mathbf{r}$ that satisfies the following equation to put an edge between atoms $i$ and $j$:
\begin{eqnarray} 
\left|\mathbf{r}-\mathbf{R}_{k}(n\Delta t)\right|>\left|\mathbf{r}-\mathbf{R}_{i}(n\Delta t)\right|
=\left|\mathbf{r}-\mathbf{R}_{j}(n\Delta t)\right|,
\label{eq:Delaunay}
\end{eqnarray}
for all other atoms $k$. Here, $\mathbf{R}_{i}(n\Delta t)$ it the location vector of atom $i$ at time $t=n\Delta t$.
For the Gabriel graph, we need to satisfy the following conditions for all other atoms $k$ to put an edge between atoms $i$ and $j$:
\begin{eqnarray} 
\left|\mathbf{R}_{k}(n\Delta t)-\frac{\mathbf{R}_{i}(n\Delta t)+\mathbf{R}_{j}(n\Delta t)}{2}\right|
\nonumber \\
>\left|\frac{\mathbf{R}_{i}(n\Delta t)-\mathbf{R}_{j}(n\Delta t)}{2}\right|.
\label{eq:Gabriel}
\end{eqnarray}
In this paper, we put Eqs.~(\ref{eq:Delaunay}) and (\ref{eq:Gabriel}) to computational algorithms in simplistic ways, with discretizing and sweeping the point $\mathbf{r}$ in the calculation of Delaunay graphs.
Figure \ref{fig:ggraph_example} shows Delaunay and Gabriel graphs for the Nd-Fe alloy containing 41~\% of Nd at a particular time step in our calculation. It is possible to see there are excessive edges of the Delaunay graph, which are indicated by red lines, compared with those of the Gabriel graph, which are indicated by black lines.

Our next task is to determine which graph gives better expression of NN structures in the amorphous Nd-Fe alloys.
To make a quantitative comparison, we defined the radial distribution functions of Delaunay and Gabriel graphs as follows:
\begin{eqnarray}
&&g^{D}_{\mathrm{AB}}(r)
\nonumber\\
&&=\frac{1}{4\pi r^2}\frac{1}{N_{\mathrm{time}}}\sum_{n=1}^{N_{\mathrm{time}}}
  \sum_{i\in\mathrm{A},j\in\mathrm{B},\langle i, j \rangle\in D(n\Delta t)}
  \!\!\!\!\!\!\!\!\!\!\!\!\delta(r-R_{ij}(n\Delta t))
\\
&&g^{G}_{\mathrm{AB}}(r)
\nonumber\\
&&=\frac{1}{4\pi r^2}\frac{1}{N_{\mathrm{time}}}\sum_{n=1}^{N_{\mathrm{time}}}
  \sum_{i\in\mathrm{A},j\in\mathrm{B},\langle i, j \rangle\in G(n\Delta t)}
  \!\!\!\!\!\!\!\!\!\!\!\!\delta(r-R_{ij}(n\Delta t)),
\end{eqnarray}
where $D(n\Delta t)$ and $G(n\Delta t)$ represent the Delaunay and Gabriel graphs at time $t=n\Delta t$ and $\langle i, j \rangle\in D(n\Delta t)$ means that an atomic connection between $i$ and $j$ is included to the edge of Delaunay graph $D(n\Delta t)$.
The summation was taken for all the edges in the graphs and averaged for all the time steps.

\begin{figure}[th]
\begin{center}
\includegraphics[width=60mm]{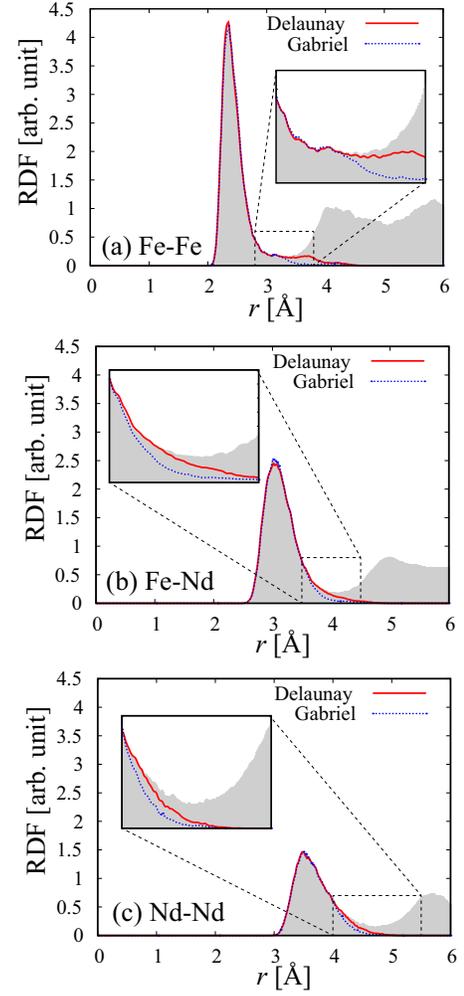}
\end{center}
\caption{
Comparison between the whole radial distribution function $g_{\mathrm{AB}}(r)$, the radial distribution function $g_{\mathrm{AB}}^{D}(r)$ of Delaunay graph and 
the radial distribution function $g_{\mathrm{AB}}^{G}(r)$ of Gabriel graph at Nd $= 41~\%$.
The figures (a), (b) and (c) show the RDFs for pairs of atomic species Fe-Fe, Fe-Nd and Nd-Nd, respectively. 
The solid red, and dotted blue lines in each figure represents RDF of Delaunay and Gabriel graphs, and the gray area represents the whole RDF. The inset in each figure shows the magnified image around the first local minimum. 
All the functions are the averages of 6 independent samples.
\label{fig:RDF_ggraph}
}
\end{figure}

Figure \ref{fig:RDF_ggraph} shows the comparison among $g_{\mathrm{AB}}(r)$, $g^{D}_{\mathrm{AB}}(r)$ and $g^{G}_{\mathrm{AB}}(r)$ for different element pairs A-B at Nd~=~41~\%. It is possible to see that both $g^{D}_{\mathrm{AB}}(r)$ and $g^{G}_{\mathrm{AB}}(r)$ give fair expressions of the first peak, and it is possible to see some interesting features in the magnified images around the first local minima.
For the Fe-Fe component, $g_{\mathrm{AB}}^{D}(r)$ has a long tail in the region of second peak, whereas $g_{\mathrm{AB}}^{G}(r)$ drops significantly around the local minimum. In contrast, $g_{\mathrm{AB}}^{G}(r)$s drop to zero before reaching the local minima for the Fe-Nd and Nd-Nd components, and $g_{\mathrm{AB}}^{D}(r)$s are slightly larger than $g_{\mathrm{AB}}^{G}(r)$s.
Given the conventional definition of first NN given by the local minima of RDFs, the features in Fig.~\ref{fig:RDF_ggraph} means that the Delaunay graph overestimates the first NNs for the Fe-Fe component, whereas the Gabriel graph underestimates the first NNs for the Fe-Nd and Nd-Nd components.

\begin{table}
\begin{center}
\begin{tabular}{c|rrrrrr}
\hline
&\multicolumn{3}{c}{$\eta^{D}_{\mathrm{AB}}$}&\multicolumn{3}{c}{$\eta^{G}_{\mathrm{AB}}$}\\
&Fe-Fe&Fe-Nd&Nd-Nd&Fe-Fe&Fe-Nd&Nd-Nd\\
\hline
20~\%&$ 13$~\%&$  2$~\%&$ -6$~\%&$  7$~\%&$  3$~\%&$-10$~\%\\
31~\%&\textcolor{red}{$ 18$~\%}&$  3$~\%&$ -3$~\%&$  8$~\%&$ -3$~\%&$ -9$~\%\\
41~\%&\textcolor{red}{$ 16$~\%}&$  4$~\%&$ -3$~\%&$  7$~\%&$ -2$~\%&$-10$~\%\\
50~\%&\textcolor{red}{$ 17$~\%}&$  5$~\%&$ -1$~\%&$  7$~\%&$ -1$~\%&$ -8$~\%\\
59~\%&\textcolor{red}{$ 21$~\%}&$  6$~\%&$  2$~\%&$  9$~\%&$  1$~\%&$ -4$~\%\\
69~\%&\textcolor{red}{$ 18$~\%}&$  7$~\%&$  3$~\%&$ 12$~\%&$  2$~\%&$ -3$~\%\\
80~\%&$ 13$~\%&$  7$~\%&$  2$~\%&$ 13$~\%&$  4$~\%&$ -3$~\%\\
\hline
\end{tabular}
\end{center}
\caption{
The over/underestimation percentages of coordination numbers $\eta^{D}_{\mathrm{AB}}$ and $\eta^{G}_{\mathrm{AB}}$ derived from the Delaunay and Gabriel graphs, respectively. The three columns in the left correspond to the different element pair components A-B of $\eta^{D}_{\mathrm{AB}}$, and the three columns in the right correspond to the different element pair components A-B of $\eta^{G}_{\mathrm{AB}}$. The percentages are indicated by red when the magnitude is larger than 15~\%.
}
\label{tab:avg_coord_num}
\end{table}

To determine which graph to be the better expression of first NNs, we compared the averaged coordination numbers determined from the Delaunay and Gabriel graphs.
The averaged coordination numbers $C^{D}_{\mathrm{AB}}$ and $C^{G}_{\mathrm{AB}}$ derived from the Delaunay and Gabriel graphs, which mean the averaged numbers of B atoms surround an A atom, can be written in the following formulae:
\begin{eqnarray}
\begin{array}{lcr}
C^{D}_{\mathrm{AB}}
&=
 &\displaystyle{
   \frac{1}{2\pi N_{\mathrm{A}}}\int dr \frac{g^{D}_{\mathrm{AB}}(r)}{r}
  }
\\
C^{G}_{\mathrm{AB}}
&=
 &\displaystyle{
   \frac{1}{2\pi N_{\mathrm{A}}}\int dr \frac{g^{G}_{\mathrm{AB}}(r)}{r},
  }
\end{array}
\end{eqnarray}
where $N_{\mathrm{A}}$ is the number of atoms of the element A.
We compared these coordination numbers with the conventional definition of averaged coordination number calculated from the whole RDF:
\begin{equation}
C_{\mathrm{AB}}
=\frac{1}{2\pi N_{\mathrm{A}}}\int_{0}^{r_{\mathrm{min}}} dr \frac{g_{\mathrm{AB}}(r)}{r},
\end{equation}
where $r_{\mathrm{min}}$ denotes the first local minima of whole RDFs.
For the quantitative comparison, we defined the over/underestimation percentages of coordination numbers as in the following formulae:
\begin{equation}
\begin{array}{lcr}
\eta^{D}_{\mathrm{AB}}
&=
 &\displaystyle{
   \left(\frac{C^{D}_{\mathrm{AB}}}{C_{\mathrm{AB}}}-1
   \right)\times100\ [\%]
  }
\\
\eta^{G}_{\mathrm{AB}}
&=
 &\displaystyle{
    \left(\frac{C^{G}_{\mathrm{AB}}}{C_{\mathrm{AB}}}-1
    \right)\times100\ [\%],
  }
\end{array}
\end{equation}
and we show these over/underestimation percentages in Table \ref{tab:avg_coord_num} for the different composition ratio.
We can see in Table \ref{tab:avg_coord_num} that the Delaunay graphs significantly overestimate the Fe-Fe components, whereas the over- or underestimation percentages of Gabriel graphs are relatively small for overall cases.

\section{Analysis of difference in orders in Nd-Fe alloys}

From the results and discussions in the previous section, we redefined the first NNs in Nd-Fe alloys by the corresponding Gabriel graphs in this paper.
Using this definition, we tried to analyze the difference in crystallinity of Nd-Fe alloys depending on the composition ratio.

\subsection{Coordination number distributions}

\begin{figure}[th]
\begin{center}
\includegraphics[width=3.5cm]{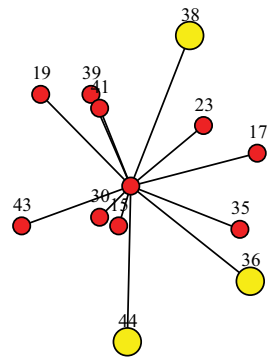}
\end{center}
\caption{
A partial Gabriel graph connected to atom 1 at a specific time in the system shown in Fig.~\ref{fig:sample}.
The large yellow circles and small red circles correspond to Nd and Fe atoms, respectively.
The numberings on the atoms represent the atomic identification numbers.
}
\label{fig:ggraph_part}
\end{figure}

By analyzing the Gabriel graph of each time step, it is possible to make further statistical analyses of the coordination numbers, not just for the averages but also for their distributions as follows.
A concept of the coordination numbers for each atom at a particular time step can be understood from the partial Gabriel graph shown in Fig.~\ref{fig:ggraph_part}.
In Fig.~\ref{fig:ggraph_part}, the number of neighboring Fe atoms around atom 1 is 9, and the number of neighboring Nd atoms around atom 1 is 3 in this case.
We regard the numbers of neighboring Fe and Nd atoms in the Gabriel graph as the coordination numbers at a particular time step.

\begin{figure*}
\begin{center}
\includegraphics[width=150mm]{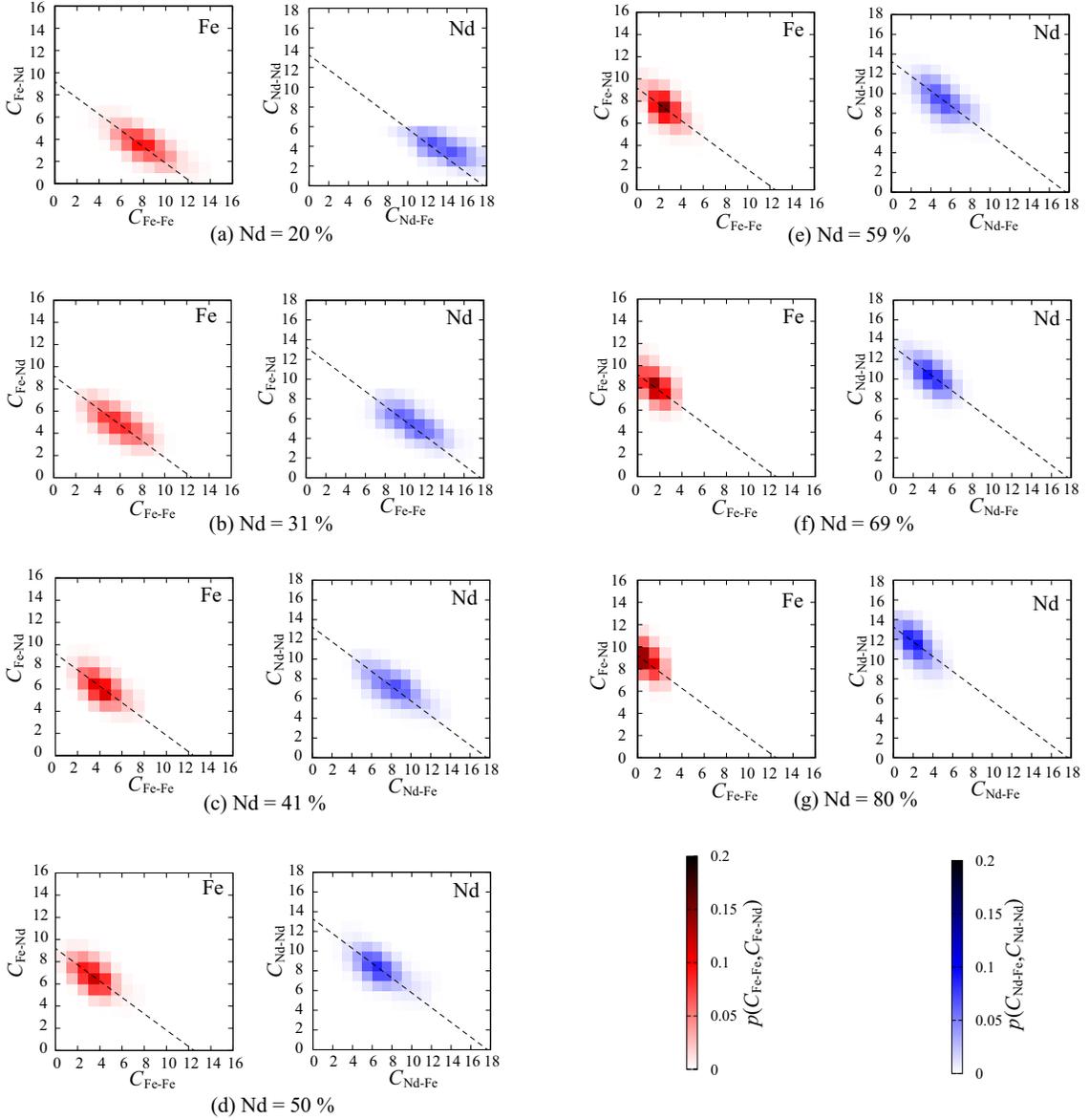}
\end{center}
\caption{
Two-dimensional distributions of the coordination numbers around an Fe atom and a Nd atom.
In each figure, the left graph corresponds to the 
probability of coordination numbers 
around Fe atoms, and the right graph corresponds to the 
probability of coordination numbers 
around Nd atoms. The histograms are the averages of 6 independent samples.
The dashed lines in the left graphs correspond to 
$C_{\mathrm{Fe\mathchar`-Nd}}+0.736\,C_{\mathrm{Fe\mathchar`-Fe}}=9.19$, 
and the dashed lines in the right graphs correspond to 
$C_{\mathrm{Nd\mathchar`-Nd}}+0.750\,C_{\mathrm{Nd\mathchar`-Fe}}=13.3$.
\label{fig:coord_num}
}
\end{figure*}

It is therefore possible to calculate the probabilities $p(C_{\mathrm{Fe\mathchar`-Fe}},C_{\mathrm{Fe\mathchar`-Nd}})$ and $p(C_{\mathrm{Nd\mathchar`-Fe}},C_{\mathrm{Nd\mathchar`-Nd}})$, which describe the tendencies when a set of coordination numbers appears the around an Fe atom and a Nd atom, respectively.
Figure \ref{fig:coord_num} shows the two-dimensional 
distributions of $p(C_{\mathrm{Fe\mathchar`-Fe}},C_{\mathrm{Fe\mathchar`-Nd}})$ and $p(C_{\mathrm{Nd\mathchar`-Fe}},C_{\mathrm{Nd\mathchar`-Nd}})$ 
at different composition ratios from Nd~=~20~\% to 80~\%.
In each figure of Fig.~\ref{fig:coord_num}, the left graph corresponds to 
$p(C_{\mathrm{Fe\mathchar`-Fe}},C_{\mathrm{Fe\mathchar`-Nd}})$
and the right graph corresponds to 
$p(C_{\mathrm{Nd\mathchar`-Fe}},C_{\mathrm{Nd\mathchar`-Nd}})$. 
A few remarkable features can be seen in Fig.~\ref{fig:coord_num}. First, the coordination numbers distribute in certain ranges for all the cases. Second, the 
high probability areas
of the coordination numbers move to the upper left gradually as the ratio of Nd increases.
This distribution shifts are approximated well by the lines of 
$C_{\mathrm{Fe\mathchar`-Nd}}+0.736\,C_{\mathrm{Fe\mathchar`-Fe}}=9.19$ 
for Fe and 
$C_{\mathrm{Nd\mathchar`-Nd}}+0.750\,C_{\mathrm{Nd\mathchar`-Fe}}=13.3$ 
for Nd, as shown by the dashed lines in Fig.~\ref{fig:coord_num}.
This means that the total coordination numbers 
($C_{\mathrm{Fe\mathchar`-Nd}}+\,C_{\mathrm{Fe\mathchar`-Fe}}$ for Fe and $C_{\mathrm{Nd\mathchar`-Nd}}+\,C_{\mathrm{Nd\mathchar`-Fe}}$ for Nd) 
decrease when increasing the Nd ratio.
Finally, the ranges of distributions of coordination numbers vary depending on the composition ratio.
These features indicate the strong dependence of coordination structures on the composition ratio in amorphous Nd-Fe alloys.

\subsection{Bond angle distribution}
\begin{figure*}
\begin{center}
\includegraphics[width=120mm]{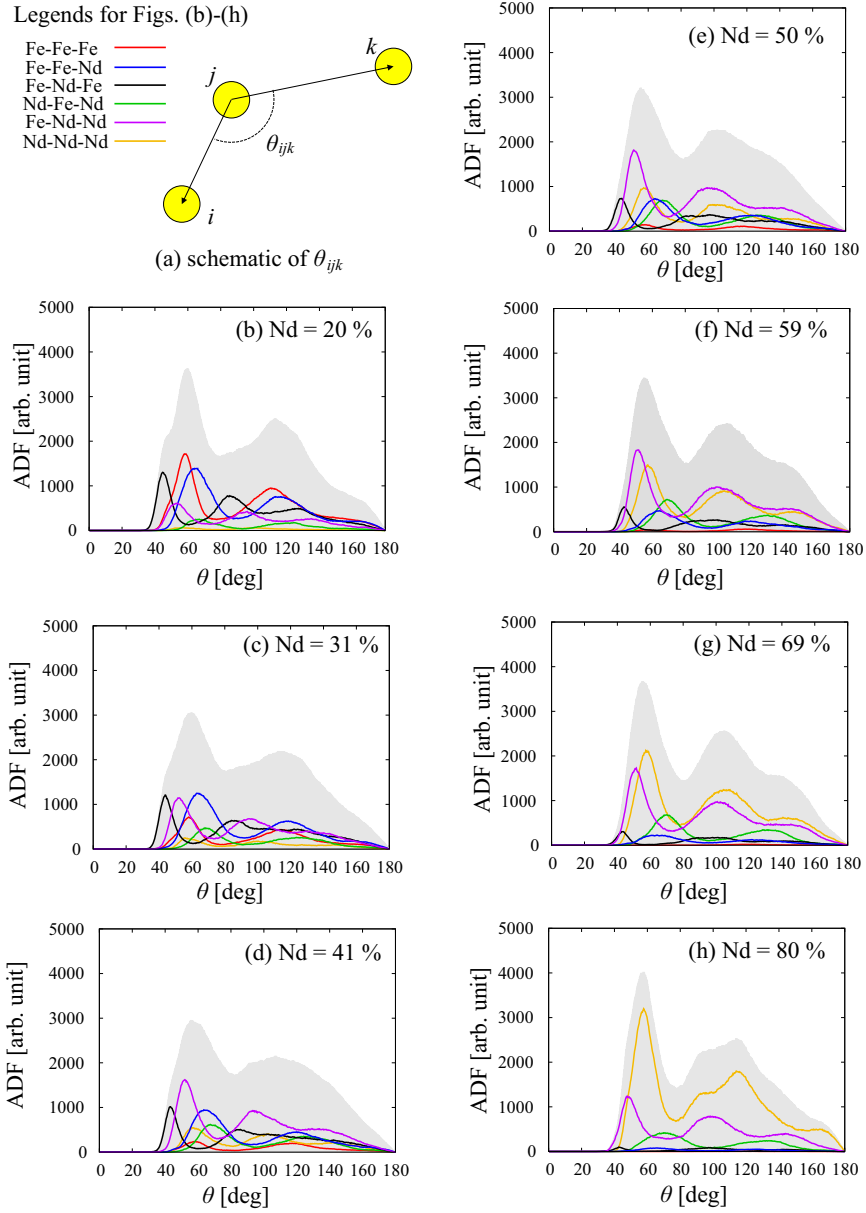}
\end{center}
\caption{(a) A schematic of the bond angle for atomic bonding $i$-$j$-$k$. (b)--(h) Angle distribution functions for the Nd-Fe alloys for composition ratios of Nd $=20~\%,~31~\%,~41~\%,~50~\%,~59~\%,~69~\%$ and $80~\%$. In each figure, the solid lines corresponds to the ADF components of different bonding structures (see Legends), and the gray area corresponds to the sum of all the components. All the functions are the averages of 6 independent samples.
\label{fig:ADF}
}
\end{figure*}

It is possible to define the bond angle $\theta_{ijk}(t)$ for a pair of adjacent edges $\mathbf{R}_{ji}(t)$ and $\mathbf{R}_{jk}(t)$ from the Gabriel graph at time $t$.
The bond angle $\theta_{ijk}(t)$ can be calculated using the atomic coordinate as follows:
\begin{equation}
\theta_{ijk}(t)
=\mathrm{arccos}\left(
\frac{\mathbf{R}_{ji}(t)\cdot\mathbf{R}_{jk}(t)}{|\mathbf{R}_{ji}(t)||\mathbf{R}_{jk}(t)|}
\right).
\label{eq:angle}
\end{equation}
(See also Fig.~\ref{fig:ADF}\,(a) for a schematic.)
From all the angles in the Gabriel graphs for all the time step, it is possible to calculate the angular distribution functions (ADFs) as follows:
\begin{eqnarray}
f_{\mathrm{ABC}}(\theta)
&=
 &\frac{1}{N_{\mathrm{time}}}\sum_{n=1}^{N_{\mathrm{time}}}\sum_{j\in\mathrm{B}}
\nonumber \\
&&\sum_{i\in\mathrm{A},\langle j,i\rangle\in G_{j}(n\Delta t)}
  \sum_{k\in\mathrm{C},\langle j,k\rangle\in G_{j}(n\Delta t)}
\nonumber \\
&&\qquad \delta (\theta-\theta_{ijk}(n\Delta t)),
\label{eq:ADF}
\end{eqnarray}
where $G_{j}(n\Delta t)$ represents the partial Gabriel graph connected to atom $j$ at time $t=n\Delta t$ (see Fig.~\ref{fig:ggraph_part} for an example), and we use $\delta(\theta) \approx (\sqrt{2\pi}\sigma)^{-1}\exp\left[ -\theta^{2}/(2\sigma ^{2})\right]$ with $\sigma=0.1~[\mathrm{deg}]$ as an approximated delta function.

Figures \ref{fig:ADF}\,(b)--(h) show the ADFs {and their summations} at different composition ratios.
We can see interesting features depending on the composition ratio of Nd.
First, a typical two-peak structures are found universally at the summations of the ADF components, which are shown as the gray areas of Fig.~\ref{fig:ADF}\,(b)--(h), whereas the steepness of the peaks are different among those graphs.
Some remarkable features can be seen depending on the composition ratio in the ADF components for the atomic conjunctions.
Below Nd $= 50~\%$, we can see mixtures of the ADF components and there are no dominant components. Above Nd $= 50~\%$, the ADF components of Nd-Nd-Fe and Nd-Nd-Nd become dominant and other ADF components are suppressed. Particularly at Nd~=~80~\%, the Nd-Nd-Nd component becomes steepest and most dominant, and it is possible to find the peaks of Nd-Nd-Nd component around 60$^\circ$, 90$^\circ$ and 120$^\circ$. This indicates that the Nd-Nd-Nd networks form closed-pack-like structures in the Nd-Fe alloys of this composition ratio.

\section{Discussion}

\subsection{Possible origin of difference in crystallinity}

\begin{figure}[th]
\begin{center}
\includegraphics[width=8cm]{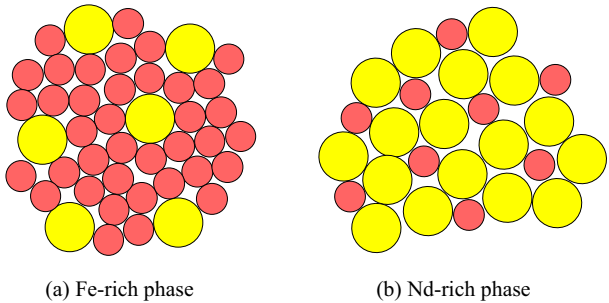}
\end{center}
\caption{
A schematic of possible atomic alignments in amorphous Nd-Fe alloys. The large yellow circles represent Nd atoms, and the small red circles represent Fe atoms.
}
\label{fig:disc}
\end{figure}

It is possible to explain some features in Fig.~\ref{fig:coord_num} by the difference between the atomic radii of Fe and Nd. 
Figure \ref{fig:disc} shows a schematic of possible atomic alignments in an Fe-rich phase and in a Nd-rich phase.
We can see in the figure that the total number of surrounding atoms around a Nd atom tends to become larger than that of an Fe atom, because the surface area of Nd is larger than that of Fe. That is the reason of the coordination numbers of Nd being larger than those of Fe and the decreases of total coordination numbers for both Fe and Nd when increasing Nd ratio.
It is also possible to explain the features of ADFs in Fig.~\ref{fig:ADF} from the atomic radii of Fe and Nd as follows.
Since the Nd radius is larger than that of Fe, Nd atoms can disturb the closed-pack-like structures of Fe in the Fe-rich phase more effectively than the opposite case. The mixture of ADF components seen in the Fe-rich cases are also originated from the large Nd radius: Nd atoms are attached to many of other atoms even when their total number is small.
In the Nd-rich phase, on the other hand, Fe atoms move to the interstitial positions of Nd structures in order to minimize the volume and therefore the closed-pack-like Nd networks appear in Nd-rich phases. This is also the reason why the coordination number $C_{\mathrm{Fe}}$ around an Fe atom becomes almost zero at Nd~=~80~\%, as seen in Fig.~\ref{fig:coord_num} (g).
These features indicate that the short-range order becomes 
stronger
in Nd-Fe alloys with the 
high 
Nd composition ratio.
Since the long-range order of crystallization may come from the those short-range orders, this could also related to the difference in crystallinity of GB phases in the experiment \cite{GB_amor}.


\section{Summary}

In this paper, we proposed a simple method to define NN networks in amorphous systems using the Gabriel graphs.
We applied this method to Nd-Fe alloys computationally created by melt-quench simulations with first-principles molecular dynamics.
We examined the Delaunay and Gabriel graphs of each composition ratio for whole time step of the finite temperature MD in order to investigate the coordination structures in statistical details.
By the comparison among the whole RDFs, the RDFs of Delaunay and Gabriel graphs for each system, it was shown that the Gabriel graph depicts the NN sites well whereas the Delaunay graph overestimates the coordination numbers for Fe-Fe components.
Using the Gabriel graphs for all the time steps, we were able to investigate the distributions of coordination numbers depending on the atomic species.
It was shown that the coordination numbers have certain ranges of distributions depending on the composition ratio.
We also examined the angular distributions calculated from the Gabriel graphs.
We found a difference in the extent of ordering depending on the Nd composition ratio, which can be related to the amorphous GB phase in the experiment \cite{GB_amor}.
Our graph analysis scheme may open up a new perspective to the complicated phenomena in amorphous systems which are not understandable in conventional approaches, like electron-related phenomena or magnetism of amorphous materials.


\begin{acknowledgments}
The authors would like to acknowledge Yasutomi Tatetsu, Kazuhiro Hono, and Satoshi Hirosawa for suggestions and comments about the GB phases in permanent magnets. We are also grateful to Yasunobu Ando and Taisuke Ozaki for extensive discussions and suggestions for overall results.

This work was supported in part by MEXT of Japan as a social and scientific priority issue CDMSI to be tackled by using post-K computer, and the Elements Strategy Initiative Project under the auspice of MEXT, as well as KAKENHI Grant No. 17K04978. 
The calculations were partly carried out by using supercomputers at ISSP, The University of Tokyo, and TSUBAME, Tokyo Institute of Technology as well as the K computer, RIKEN (Project Nos.~hp170269 and hp180206).
\end{acknowledgments}

\end{document}